\documentstyle [11pt]{article}
\newcommand{\sect}[1]{\setcounter{equation}{0}\section{#1}}

\newcommand{\f}{\frac}
\newcommand{\r}{\rho}
\newcommand{\br}{\bar{\r}}

\newcommand{\p}{\partial}

\newcommand{\bp}{\bar{\partial}}

\newcommand{\s}{\star}

\newcommand{\1}{{\bf 1}}

\newcommand{\ot}{\otimes}
\newcommand{\bphi}{\bar{\phi}}
\newcommand{\Vphi}{\large{\varphi}}
\newcommand{\La}{\Lambda}
\newcommand{\la}{\lambda}
\newcommand{\tla}{\tilde{\lambda}}
\newcommand{\A}{{\cal A}}
\newcommand{\h}{{\cal H}}

\newcommand{\ap}{\approx}
\newcommand{\bc}{\begin{center}}
\newcommand{\ec}{\end{center}}
\newcommand{\be}{\begin{equation}}
\newcommand{\ee}{\end{equation}}

\newcommand{\uot}{\underline{\ot}}

\def\lcross{{>\!\!\!\triangleleft}}

\newcommand{\cn}{{\bf C}}
\newcommand{\rn}{{\bf R}}
\newcommand{\zn}{{\bf Z}}
\newcommand{\nn}{{\bf N}}
\newtheorem{prop}{Proposition}
\newtheorem{lemma}{Lemma}

\begin{document}

\bc
{\huge{\bf Many q-Particles from One: a New}}
\ec
\bc
{\huge{\bf Approach to $*$-Hopf Algebras?}}
\ec

{}~
\bc
{\bf Gaetano Fiore\footnote{Alexander-von-Humboldt fellow}}
\ec
\bc
{\it Sektion Physik der Universit\"at M\"unchen, Ls. Prof. Wess}
\ec
\bc
{\it Theresienstrasse 37, D-80333 M\"unchen, Germany}
\ec
\bc
LMU-TPW 94-15, hep-th/9411006, October 1994
\ec

\section*{\center Abstract}
We propose a nonstandard approach to solving the apparent
incompatibility between
the coalgebra structure of some inhomogeneous quantum groups and their natural
complex conjugation. In this work we sketch the general idea and develop the
method in detail on a toy-model; the latter is a q-deformation of the
Hopf algebra of 1-dim translations + dilatations. We show how to get all
Hilbert space representations of the latter from tensor products of
the fundamental ones; physically, this corresponds to constructing
composite systems of many free distinct q-particles in terms of the basic
one-particle ones. The spectrum of the total momentum turns out to be the
same as that of a one-particle momentum, i.e. of the form
$\{\mu q^n\}_{n\in\zn}$.

\vskip1truecm

\sect{Introduction and preliminaries}

{}~~~The problem at the root of this work is the well-known
apparent incompatibility
between the Hopf structure of some quantum groups and their natural
$*$-structure (i.e. complex
conjugation). This incompatibility undermines the possibility of
constructing composite quantum-mechanical systems with the corresponding
q-group symmetry in terms of elementary ones. We have in mind especially
q-deformations of inhomogenous Lie group
such as Poincare's and the Euclidean one (which in the `` undeformed ''
physics play an essential role as fundamental space(time) symmetries)
in the form of braided semidirect products
$IG_q:=S_q\lcross G_q$, \cite{schl,maj2,wess,fioeu}; here $G_q$ denotes an
homogeneous
quantum group and $S_q$ the corresponding quantum space. When $q\in\rn^+$
$Fun(IG_q)$ is a $*$-coalgebra but not a $*$-algebra, its dual
$U_q(ig)$ is $*$-algebra but not a $*$-coalgebra
(i.e. $(*\ot *)\circ \phi\neq \phi \circ *$). When dealing with
$*$-representations $\r$ of $U_q(ig)$ on Hilbert spaces, this implies
in particular that if $\r(O)\in \r(U_q(ig))$ is an observable of a simple
system, $(\r\ot \r)\circ\phi(O)$ in general is {\it not} an observable of the
composite system, since it is not hermitean.

The typical form of the coproduct of $U_q(ig)$ is
\be
\phi(u)=u_i\ot u'_i~~~~~~~~~~~~~~~\phi(p^i)=p^i\ot \1+\la u^i_j\ot p^j
{}~~~~~~~~~~~~~~~~u,u_i,u'_i,u^i_j\in U_q(g);
\ee
its restriction to the homogeneous sub-Hopf algebra $U_q(g)$ commutes with
$*$, but $\phi,*$ don't commute on the translation generators $p^i$,
due to the presence of the `` dilaton '' $\la$:
\be
\la p^i=q^{-1}p^i\la,~~~~~~~~~~~~[\la,u]=0,~~~~~~~~~~u\in U_q(g);
\ee
in fact, $(p^i)^*=p^jC_{ji}$
($C$ is a matrix of numbers) and $\la^*=\alpha\la^{-1}$ (with some
$\alpha \in \cn$). Nevertheless, $*$ maps the above coalgebra structure
into a `` conjugated '' one, which is also compatible with the algebra
structure of $U_q(ig)$.

The latter observation is at the hearth of our idea
of using both `` conjugated '' coalgebras to form a genuine
$*$-Hopf algebra isomorphic to $U_q(ig)$, and of how to use the latter
in $*$-representation theory. Here we illustrate the idea in a most
simple and pedagogical toy-model, representing a q-deformation of the
abelian algebra of 1-dim translations (enlarged with dilatations)
and of its Hilbert space representations. The spurious dilatation generator
$\la$ (which is absent in the undeformed
Hopf algebra) is introduced to mimic the features of the Hopf-algebras
$U_q(ig)$ summarized above. The algebra can be considered as the
appropriate symmetry of some suitable q-deformed quantum mechanical
systems consisting of one (or many) free particles on the quantum line $\rn_q$.
In a forthcoming paper \cite{fionew}
we are going to deal with the general problem sketched
above and make contact with the work in Ref. \cite{maj1}.

We will assume that the reader is familiar with the axioms of
bialgebras/ Hopf-algebras. Given a coassociative coproduct
$\phi:\A\rightarrow \A\ot\A$,
we will denote by $\phi_n:\A\rightarrow\bigotimes^n\A$
the map which can be obtained recursively in one of the following ways
\be
\phi_{n+1}=\phi_{n,j}\phi_n,~~~~~~~~~~~~~~~~~
\phi_{n,j}:=(\underbrace{id\ot...\ot id}_{(n-j)~~times}\ot\phi\ot
\underbrace{id\ot...\ot id}_{(j-1)~~times})~~~~~~~~~~1\le j\le n.
\ee
It will be useful to introduce the function
\be
(a;q)_n:=\prod\limits_{i=0}^{n-1}(1-aq^i),~~~~~~~~~~~~~a\in\cn,~~~n\in \nn;
\ee
when $|q|<1$ we will define
$(a;q)_{\infty}:=\lim_{n\rightarrow \infty}(a;q)_n$.

\sect{The toy-model}

The building block of our toy-model
consists of one free quantum particle on a quantum line
$\rn_q$, $q\in\rn^+$, as considered in Ref. \cite{heb}.

The starting algebra $\A$ is the unital $*$-algebra generated by elements
$\{p^{\pm 1},\la^{\pm 1}\}$ (actually we will assume that it is possible
to extend it also to rational functions of the generators)
satisfying the algebra relations
\be
p\la=q\la p,~~~~~~\la^{\mp 1}\la^{\pm 1}=\1=p^{\mp 1}p^{\pm 1},
\ee
and endowed with an antilinear involutive antihomomorphism $*$
(the `` complex conjugation '') such that on the generators
\be
(p^{\pm 1})^*=p^{\pm 1},~~~~~(\la^{\pm 1})^*=\la^{\mp 1}.
\ee
$p$ plays the role of q-deformed translation generator in 1-dim configuration
space; $\la$ of q-deformed dilatation generator, or equivalently
(since the model is one-dimensional), of q-deformed generator of boosts.

One can introduce a $*$-irrep (i.e. a $*$-irreducible representation) of $\A$
on a Hilbert space $\h_{\mu}$ in the following way. One postulates
that $*$ also represents the operation of hermitean conjugation
of operators and introduces an orthonormal basis of $\h_{\mu}$
$\{|n>\}_{n\in\zn}$ consisting of eigenfunctions of $p$:
\be
p|n>=q^n\mu|n>,
\ee
where $\mu$ is a constant with dimension of a mass;
$p$ plays the role of q-deformed momentum observable. $\la$ is  a unitary
`` step '' operator mapping eigenfunctions of $p$ into each other:
\be
\la^{\pm 1}|n>=|n\pm 1>.
\ee
For the sake of simplicity,
here and in the sequel we will use the same symbols $p,\la$ to denote
both the abstract elements of $\A$ and their representations as Hilbert space
operators on $\h_{\mu}$.
Each vector $|n>$ is cyclic in $\h_{\mu}$, since $|m>=\la^{m-n}|n>$.
Note that the state of rest $p=0$ does not belong to $\h_{\mu}$
(it belongs to some extension of
$\h_{\mu}$, i.e. to the space of functionals on some dense domain
${\cal S}\subset \h_{\mu}$ of vectors `` rapidly decreasing '' at $\infty$),
but can be approached as much as desired.

One can extend $\A$ to a q-deformed Heisenberg algebra $\A_H$ \cite{heb}
by the introduction
of a formally hermitean q-deformed position operator $x$ defined \cite{heb}
as a function of $p,\la$ (see also Ref. \cite{schwenk});
a Hilbert space representation of $\A_H$ cannot be found on $\h_{\mu}$,
but it can be found  on its direct sum with the Hilbert space of opposite
momentum, $\h_{\mu}\oplus\h_{-\mu}$. For the moment we will consider only
irreps of $\A$, i.e. systems with momentum having a well defined sign.

Next, we choose the Hopf algebra structure $\{\phi,\epsilon,S\}$
for our toy model; this is needed
to construct many-particle systems by using one particle
ones as building blocks.

One possibility is the classical (in the sense of undeformed) one.
The actions of $\phi_c,\epsilon_c,\sigma_c$ on the
generators have the form
\be
\phi_c(p)=p\ot \1+\1\ot p~~~~~~~\phi_c(\la^{\pm 1})=\la^{\pm 1}\ot\la^{\pm 1}
{}~~~~~~~~\phi_c(\1)=\1\ot\1
\ee
\be
\epsilon_c(p)=0~~~~~~\epsilon_c(\la^{\pm 1})=1=\epsilon_c(\1)
\ee
\be
\sigma_c(p)=-p~~~~~~~\sigma_c(\la^{\pm 1})=\la^{\mp 1}~~~~~~~~~\sigma_c(\1)=\1,
\ee
and are extended to all of $\A$ as algebra homomorphisms
($\phi,\epsilon$) or antihomomorphism ($\sigma$) respectively.
The physical meaning of any Hopf structure $\{\phi,\epsilon,\sigma\}$ for
$\A$ should be
the following. $\phi$ should allow to represent the algebra relations (2.1)
on the Hilbert space $\h$ of states of a system composed of two subsystems;
$\h$ should be built as the tensor
product of the Hilbert spaces $\h_1,\h_2$ of the two subsystems. For instance,
$\phi(p)$ should be an observable and should represent the momentum operator
of the composite system, $\phi(\La)$ the corresponding step operator.
$\epsilon(p),\epsilon(\La)$ should give the eigenvalues of the operators
$p,\La$ on the vector with $p=0$.
 $\sigma(p)$ should represent the `` opposite '' momentum operator of $p$
w.r.t. the `` sum '' operation introduced by $\phi$, in the sense that
$(id\ot \sigma)\circ\phi(p)$ and
$(\sigma\ot id)\circ\phi(p)$) should represent the relative momenta of
each subsystem in the rest frame of the other; the constraint
$m\circ(id\ot \sigma)\circ\phi(p)=0$ ($m(a\ot b):=ab$), which is a consequence
of the axioms of an Hopf algebra and of definition $\epsilon(p)=0$,
would guarantee
that the relative momentum of one system w.r.t. its own rest frame is
zero.

The Hopf structure (2.5-7)
is compatible with the definition (2.2) of $*$ in the sense that
$(*\ot *)\circ\phi_c=\phi_c\circ *$ etc. It is immediate to verify that the
spectrum of $\phi(p)$ on the tensor product of two Hilbert spaces
$\h_{\mu_a}\ot\h_{\mu_b}$
$doesn't$ coincide with the spectrum of $p$ either on $\h_{\mu_a}$, or
on $\h_{\mu_b}$.
However, the definition of $\phi_c$ automatically fulfills
the physical requirement that in any normalized eigenvector
$|\psi>_{1\ot 2}=|n>_1\ot|m>_2$ of $\phi_c(p)$, $\1\ot p$
\be
|<\psi|\phi(p)|\psi>_{1\ot 2}|~>~\cases{|<\psi|\1\ot p|\psi>_{1\ot 2}| \cr
|<\psi|p\ot\1|\psi>_{1\ot 2}|\cr}~~~~~~~~~~~if~~\f{\mu_a}{\mu_b}>0
\ee
namely the the total momentum is greater than the modulus of the
momentum of either subsystem,  if
they have momenta with the same sign.

We are rather interested in a different Hopf structure, which more closely
mimics the ones of actual inhomogeneous quantum groups $U_q(ig)$,
and essentially presents the same kind of incompatibility with the
$standard$ $*$-structure. The corresponding $\phi,\epsilon,S$ act on the
generators according to the relations
\be
\phi(p)=p\ot \1+\la\ot p~~~~~~~\phi(\la^{\pm 1})=\la^{\pm 1}\ot\la^{\pm 1}
{}~~~~~~~~\phi(\1)=\1\ot\1
\ee
\be
\epsilon(p)=0~~~~~~\epsilon(\la^{\pm})=1=\epsilon(\1)
\ee
\be
\sigma(p)=-\la^{-1}p~~~~~~~\sigma(\la^{\pm 1})=\la^{\mp 1}~~~~~~~~~
\sigma(\1)=\1,
\ee
and is extended to functions of $p,\la$ in the same standard way as before.
The above is not compatible with the standard $*$-structure, since
$(*\ot *)\circ\phi\neq\phi\circ *$. Nevertheless, it satisfies the
following nonstandard properties
\be
(*\ot *)\circ\phi=\bar\phi\circ *,~~~~~*\circ\epsilon=\bar{\epsilon}\circ *
{}~~~~~~~~~~\sigma\circ *\circ \bar{\sigma}\circ *=id
\ee
Here $\{\bar{\phi},\bar{\epsilon},\bar{\sigma}\}$
is the Hopf structure which is obtained
by the replacement $\la\rightarrow \la^{-1}$ in both sides of
equations (2.5-7), i.e.:
\be
\bar{\phi}=p\ot \1+\la^{-1}\ot p~~~~~~~
\bar{\phi}(\la^{\pm 1})=\la^{\pm 1}\ot\la^{\pm 1}
{}~~~~~~~~\bar{\phi}(\1)=\1\ot\1
\ee
\be
\bar{\epsilon}(p)=0~~~~~~\bar{\epsilon}(\la^{\pm 1})=1=\bar{\epsilon}(\1)
\ee
\be
\bar{\sigma}(p)=-\la p~~~~~~~\bar{\sigma}(\la^{\pm 1})=\la^{\mp 1}
{}~~~~~~~~~\bar{\sigma}(\1)=\1,
\ee
(note that $\bar{\epsilon}=\epsilon$).

The two Hopf structures $\{\phi,\epsilon,\sigma\}$,
$\{\bar{\phi},\bar{\epsilon},\bar{\sigma}\}$
are the bosonizations of the two basic braided
Hopf structures associated to the quantum line $\rn_q$ (making it
a ``  braided line '' \cite{kempf}),
with braiding respectively given by
\be
\psi(p\uot p)=q^{-1} p\uot p~~~~~~~~~~\bar{\psi}(p\uot p)=qp\uot p.
\ee
An analogous pair of conjugated (braided) Hopf algebras arises also
for quantum spaces which are comodule algebras of a bialgebra $A(R)$
\cite{frt}, where $R$ is a solution of the Yang-Baxter equation
(see Ref. \cite{fionew}).

\section{The nonstandard $*$-realization of the Hopf algebra
$\{\A,\phi,\epsilon,\sigma\}$}

It is our (philosophical) viewpoint that the barred and unbarred
Hopf structures (or equivalently
their braided versions) should be considered as
two faces of the same medal, since the underlying homogenous q-group $G_q$
symmetry is the same, and we hope to report soon on this point at a
more abstract level elsewhere
\cite{fionew}.

Here we are going to show in the case of the toy
model how to construct from their $pair$
an actual $*$-Hopf algebra, and how the latter should be used in
representation theory on Hilbert spaces; as an Hopf algebra, it
is isomorphic to
either the unbarred or the  barred version. Our arguments will be
rather heuristic in order to motivate the construction.

To implement the abovementioned viewpoint we start from the trivial
observation that
$\gamma:=\bphi\circ\phi^{-1}$ is an algebra isomorphism of
$\phi(\A)$ onto $\bphi(\A)$ with the property
\be
(*\ot *)\circ\gamma=\gamma^{-1}\circ(*\ot *),
\ee
due to equation $(2.12)_1$; moreover if we define $\Phi(a)$ as the $pair$
$\Phi(a):=(\phi(a),\bphi(a))$ and define a multiplication
$\Phi(a)\Phi(b):=\Phi(ab)$, the set $\Phi(\A)$ gets the same algebra
structure as $\A$, $\Phi(A)\ap\A$. Then setting
\be
*_2:=\tau\circ\left((*\ot *),(*\ot *)\right)
\ee
defines an antilinear involutive antihomomorphism on
$\Phi(\A)$, as a consequence of property (2.12); here $\tau[(A,B)]:=(B,A)$.
$*_2$ commutes with $\Phi$, in the sense that
$*_2\circ\Phi=\Phi\circ *$; as a consequence, $\Phi$ maps real (w.r.t.
$*_2$) elements of $\A$ into real elements of $\Phi(A)$.

This suggests that we look for an extension of the algebra isomorphism
$\gamma$ to all of $\A\ot\A$ preserving relation (3.18),
so that $*_2$ can be
extended to an antilinear involutive antihomomorphism of the whole algebra
$(\A\ot\A)_d:=\{(\alpha,\gamma(\alpha))~~|~~\alpha\in\A\ot\A\}$ onto itself.
Here the multiplication of$(\A\ot\A)_d$ is defined by
$(\alpha,\gamma(\alpha))\cdot(\beta,\gamma(\beta)):=
(\alpha\cdot\beta,\gamma(\alpha\cdot\beta))$, so that
$(\A\ot\A)_d\ap \A\ot\A)$.  Given an element
$A\in(\A\ot\A)_d$, we will introduce its decomposition in unbarred and
barred components by $A=:(\r_2(A),\br_2(A))$; consequently,
$\gamma=\br_2\circ\r_2^{-1}$ and $(*\ot *)\circ\r_2=\br_2\circ(*\ot*)$.

If this is achieved,
then, we can reiterate the construction for higher tensor products,
 defining $\gamma_n:=\bphi_n\circ(\phi_n)^{-1}$,
$\Phi_n(a):=(\phi_n(a),\bphi_n(a))$
and antilinear involutive antihomomorphisms $*_n$ ($n\ge 1$) through
\be
*_n:=(\r_n(*_n),\br_n(*_n)):=\tau\circ \left(\bigotimes^n *,\bigotimes^n *
\right),
\ee
($\phi_2\equiv \phi$,  $\bphi_2\equiv \bphi$,
$\gamma_2\equiv\gamma$, $*_1\equiv *$, $\gamma_1\equiv id$)
first on $\Phi_n(A)$, then
by an extension of $\gamma_n$ possibly on the whole
$(\bigotimes^n\A)_d:=\{(a,\gamma_n(a))~~|~~a\in\bigotimes^n\A\}\ap
\bigotimes^n\A$ (the multiplication in $(\bigotimes^n\A)_d$ is defined by
$(a,\gamma_n(a))\cdot(b,\gamma_n(b)):=
(a\cdot b,\gamma(a\cdot b))$), satisfying the fundamental
property that it commutes with $\Phi_n:=(\phi_n,\bphi_n)$, in the
sense that $*_n\circ\Phi_n=\Phi_{n-1}\circ *_{n-1}$.
If now one defines $E:=(\epsilon,\bar{\epsilon})$,
$S:=(\sigma,\bar{\sigma})$, and {\Large $*$}$|_{(\bigotimes^n\A)_d}:=*_n$,
then it is straightforward to verify that
$\{\A_d,\Phi,E,S\}$ is a Hopf algebra isomorphic to
$\{\A,\phi,\epsilon,\sigma\}$ and, if equipped with {\Large $*$},
satisfies the axioms of
a  $*$-Hopf algebra; we will call it a $nonstandard$ $*$-realization of
$\{\A,\phi,\epsilon,\sigma\}$.

Note that definition (3.20) is equivalent to
\be
\cases{\r_n(*_n):=\gamma_n^{-1}\circ\left(\bigotimes^n *\right) \cr
\br_n(*_n):=\gamma_n\circ\left(\bigotimes^n *\right) \cr}
\ee
and one can give to the above structure a more conventional form
by noting that $\A$ endowed with either $\{\phi,\epsilon,\sigma,\r(*)\}$ or
$\{\bphi,\bar{\epsilon},\bar{\sigma},\br(*)\}$ forms a $*$ Hopf-algebra.
Nevertheless, we prefer to present it as shown, since at the
representation-theoretic level both the first and the second component
of $(\bigotimes^n\A)_d$ will be needed to act on two different
vector-space realizations of the same Hilbert space (see section 4).

In section 4 we will clarify
the relevance for representation theory of $nonstandard$ realization of a
$*$-Hopf algebra.
Next, let us show that such a structure can
be realized in a rather simple way in our case.

For the sake of brevity, let us set
$$
P_2:=\Phi(p),~~~~~~~~~~
\La_2:=\left((\la p^{-1}\ot\1)\phi(p),(\la p\ot\1)\f1{\bphi(p)}\right),
$$
\be
P_1:=(\1\ot p,\1\ot p)~~~~~~~~~~\La_1:=\left(\f1{\phi(p)}(p\ot\la),
\bphi(p)(p^{-1}\ot\la)\right);
\ee
note that $\Phi(\La)=\La_1\La_2$.  Morally, $P_2$ would be the total momentum
of two particles, $P_1$ the momentum of one particle (the one
represented in the second tensor factor), $\La_i$ the corresponding
step operators.
It is easy to check that
$P_1,\La_1,P_2,\La_2$ have well-defined commutation relations
$$
[P_1,\La_1]_q=0=[P_2,\La_2]_q~~~~~~~[\La_1,\La_2]=0
$$
\be
[P_1,P_2]=0~~~~~~~[P_1,\La_2]=0~~~~~~~~[P_2,\La_1]=0,
\ee
i.e. that both the $\r_2$ and $\br_2$ images of these (q)-commutators
 actually vanish as a consequence of relations (2.1). Moreover,
$\r_2(P_i^{\pm 1}),\r_2(\La_i^{\pm 1})$ ($i=1,2$) form a set of generators of
$\A\ot\A$, and so do $\br_2(P_i^{\pm 1}),\br_2(\La_i^{\pm 1})$. Using
definition (3.20) we arrive at the complex conjugation $*_2$
\be
P_2^{*_2}=P_2~~~~~~~P_1^{*_2}=P_1~~~~~~~~~\La_1^{*_2}=\La_1^{-1}~~~~~~~~~~
\La_2^{*_2}=\La_2^{-1}
\ee
which is compatible with the algebra structure (3.23). In particular,
we see that $P_2,P_1$ are real and make up a Cartan subalgebra of
$(\A\ot\A)_d$.

One can now ask whether the above is the unique possible
$nonstandard$ $*$-realization of
$\{\A,\phi,\epsilon,\sigma\}$. This amounts to investigating the uniqueness of
$\gamma$. $\gamma$ as an isomorphism with the abovementioned properties
is not unique; but it is the only one
which can be represented as an isomorphism of operators defined within the
Hilbert space $\h_{\mu_a}\ot\h_{\mu_b}$ (see the appendix), namely the only
one which can be used for physical purposes. Therefore,
the unique complete set of commuting observables (including $P_2$) for
a 2-particle system is $\{P_2,P_1\}$, and there is no ambiguity
in the physics which can be drawn from the above scheme.

\section{Hilbert space representation of the $*$-algebra
$\{(\A\ot \A)_d\}$}

{}~~~The rationale behind nostandard $*$-realizations of
Hopf algebras (resp. of bialgebras) is that they can be used to find
Hilbert space representations of them.
The underlying idea is the following. Since an element $A\in(\A\ot \A)_d$
can be realized both by $\r(A)$ and $\br(A)$, we look for an unbarred and a
barred carrier space
$\r({\cal U}),\br({\cal U})\subset\h_{\mu_a}\ot\h_{\mu_b}$
for $\r(A),\br(A)$ respectively to act upon.
 Of course, there should be a
one-to-one correspondence between the unbarred and barred carrier spaces,
in such a way that all physical quantities (eigenvalues of observables)
are preserved under this mapping. The tricky point is that the barred
and unbarred realization are used simultaneously in defining the
`` right '' scalar product, namely the one for which hermitean conjugation
is a realization of $*_2$.

 Let $\h_{\mu_a},\h_{\mu_b}$ be the carrier spaces
of two Hilbert space representations of $\A$.
We will consider the case when the subsystems
consist of distinct particles and use a classical statistics;
consequently the result will be different according to which particle we
put in the first, second,... factor in the tensor product. We will deal with
identical particles elsewhere. We look for
$\r({\cal U}),\br({\cal U})\subset\h_{\mu_a}\ot\h_{\mu_b}$ and a
vector space isomorphism
$\gamma:\r({\cal U})\rightarrow\br({\cal U})$
such that
\be
\gamma\left[\r(A)|\psi>\right]=\gamma[\r(A)]\gamma(|\psi>)\equiv
\br(A)\gamma(|\psi>),~~~~~~~~~~~~~A\in(\A\ot \A)_d,~~~~|\psi>\in
\r({\cal U}).
\ee

Let ${\cal C}$ be a real Cartan subalgebra of $(\A\ot\A)_d$.
Assume we have determined eigenvectors
$\vert \varphi>,\vert\bar{\varphi}>\in \h_{\mu_a}\ot \h_{\mu_b}$
of $\r_2({\cal C}),\br_2({\cal C})$ respectively, with the
{\it same weight}. Define the vector space
\be
{\cal U}:=\{(\r_2(\alpha)\vert \varphi>,\br_2(\alpha)\vert\bar{\varphi}>)~~|~~
\alpha\in(\A\ot\A)_d\}\subset(\h_{\mu_a}\ot \h_{\mu_b},\h_{\mu_a}\ot
\h_{\mu_b}).
\ee
Let $<~~\Vert~~>_{a\ot b}$  be the
ordinary scalar product in $\h_{\mu_a}\ot\h_{\mu_b}$, i.e.
if $\Vert f>=|f'>_a\ot|f''>_b$, $\Vert g>=|g'>_a\ot|g''>_b$ then
$<f\Vert g>_{a\ot b}:=<f'|g'>_a<f''|g''>_b$. It has the
property that
\be
<f\Vert u~g>_{a\ot b}=<u^{*\ot *}f\Vert g>_{a\ot b},~~~~~~~~~~u\in\A\ot\A.
\ee
\begin{lemma}
The formula
\be
<U\Vert V>:=<\bar u\Vert v>_{a\ot b}+<u\Vert \bar v>_{a\ot b}~~~~~~~~~
{}~~~~~~~~~~~~~\cases{
\Vert U>\equiv (\Vert u>,\Vert\bar u>)\in{\cal U} \cr \Vert V>\equiv (\vert v>
,\Vert\bar v>)\in{\cal U} \cr}
\ee
defines a (formal) sesquilinear inner product in ${\cal U}$ such
that
\be
<U\Vert V>^*=<V\Vert U>,
\ee
and a $*_2$-representation of $(\A\ot\A)_d$ on ${\cal U}$.
\end{lemma}
$Proof$ . Sesquilinearity $<a_iU_i,b_jV_j>=a_i^*b_j<U_i,V_j>$
($a_i,b_j\in\cn$) is trivial, and so is property (4.29). The easy but
fundamental point is to show that hermitean conjugation w.r.t. the scalar
product (4.28) is a realization of $*_2$ of $(\A\ot\A)_d$.
To this end note that
$$
<U\Vert AV>-<A^{*_2}U\Vert V>\stackrel{(4.28),(3.19)}{=}
<\bar u\Vert \r(A)v>_{a\ot b}-<[\r(A)]^{*\ot *}\bar u\Vert v>_{a\ot b}
$$
\be
+<u\Vert\br(A) \bar v>_{a\ot b}-<[\br(A)]^{*\ot *}u\Vert \bar v>_{a\ot b}
\stackrel{(4.27)}{=}0.~~~~~~~~~~~~~~~~~~~\diamondsuit
\ee

The above definition is formal, i.e. it makes sense only if the RHS
of equation (4.28) (where an infinite sum can appear) is finite. This can
be always achieved by suitably shrinking ${\cal U}$. However,
we will impose on ${\cal U}$ a stronger requirement
(which we have already stated in the inclusion relation in
eq. (4.26)), namely that
for each element $\Vert U>\equiv (\Vert u>,\Vert\bar u>)$
both $\Vert u>$ and $\Vert\bar u>$
have finite norm in $\h_{\mu_a}\ot\h_{\mu_b}$ (it is immediate to
verify that this implies that $\Vert U>$ has a finite norm w.r.t. the
inner product (4.28): it follows from the positivity and finiteness of
the norm of $\Vert u>-\Vert\bar u>$ in
$\h_{\mu_a}\ot\h_{\mu_b}$).
This serves for $\gamma:\r({\cal U})\rightarrow\br({\cal U})$ to map a
subspace of $\h_{\mu_a}\ot\h_{\mu_b}$ into a subspace of
$\h_{\mu_a}\ot\h_{\mu_b}$.

The last requirement that inner product (4.28) must fulfill, in order
that it may be taken as a scalar product in ${\cal U}$, is its positivity.
It will be proved below (Proposition 1).

{\bf Remark}. Definition (4.28) has its source of inspiration in Ref.
\cite{fio}, see also section 6.

{}~~

Now let us enforce this general idea to our toy  model.

{\bf Remark}. From now on we need to specify whether $q$ is greater or smaller
than 1. We will explicitly consider only the case  $0<q< 1$.

We choose ${\cal C}$ as the subalgebra generated by $P_1,P_2$; since
both $\r(P_1)=\1\ot p$ and $\br(P_1)=\1\ot p$
act only on the second tensor factor of
$\h_{\mu_a}\ot \h_{\mu_b}$, the eigenvectors
$\Vert \varphi>,\Vert\bar{\varphi}>\in \h_{\mu_a}\ot \h_{\mu_b}$
are to be searched in the form
\be
\Vert \varphi_{n_1}>=|\varphi_{n_1}>_a\ot|n_1>_b,~~~~~~~~~~
\Vert\bar{\varphi}_{n_1}>=|\bar{\varphi}_{n_1}>_a\ot|n_1>_b
\ee
with
\be
\cases{
|\varphi_{n_1}>_a=\sum\limits_{n=-\infty}^{\infty}\varphi_{n,n_1}|n>_a  \cr
|\bar{\varphi}_{n_1}>_a=\sum\limits_{n=-\infty}^{\infty}\bar{\varphi}_{n,n_1}
|n>_a  \cr}
\ee
Consequently, $P_1\Vert\Vphi_{n_1}>=\mu_bq^{n_1}\Vert\Vphi_{n_1}>.$
We can assume without loss of generality that
$q\le\vert\f{\mu_b}{\mu_a}\vert<1$, because this can be always
achieved after a shift of the integers $l$ which label
the states $|l>_i$ of either $\h_{\mu_a}$ or $\h_{\mu_b}$.

Let $\mu$ be the would-be eigenvalue in the equation
$P_2\Vert\Vphi_{n_1}>=\mu\Vert\Vphi_{n_1}>$; then in the unbarred
representation
the latter explicitly reads
\be
\left[\sum\limits_{n=-\infty}^{\infty}\varphi_{n,n_1}(\mu_aq^n-\mu)|n>_a+
\varphi_{n,n_1}\mu_b q^m|n+1>_a\right]\ot|n_1>_b=0
\ee
implying recursive relations
\be
\varphi_{n,n_1}(\mu_aq^n-\mu)+\varphi_{n-1,n_1}\mu_bq^{n_1}=0
\ee
for the coefficients $\varphi_{n,n_1}$. Similarly for the coefficients
$\bar{\varphi}_{n,n_1}$ we find the relations
\be
\bar{\varphi}_{n,n_1}(\mu_aq^n-\mu)+\bar{\varphi}_{n+1,n_1}\mu_bq^{n_1}=0.
\ee
Let us see investigate for which values of $\mu$ the vector
$\Vert\Vphi_{n_1}>$ has a positive and finite norm w.r.t. the
product formally introduced in equation (4.28). In terms of the
coefficients $\varphi_{n,n_1}$, $\bar{\varphi}_{n,n_1}$ this would-be
norm reads
\be
<\Vphi_{n_1}\Vert\Vphi_{n_1}>=\sum\limits_{n=-\infty}^{\infty}
\bar{\varphi}_{n,n_1}^*\varphi_{n,n_1}~~+~~c.~c.
\ee
It is easy to realize that convergence of the above series for large
$n$ requires the existence of a $n_2\in{\bf Z}$ such that $\mu=\mu_aq^{n_2}$,
because otherwise
\be
\cases{
\varphi_{n,n_1}=\varphi_{s,n_1}(\f{\mu_bq^{n_1}}{\mu})^{n-s}
\f1{(\f{\mu_a}{\mu}q^{s+1};q)_{n-s}} \cr
\bar{\varphi}_{n,n_1}=\bar{\varphi}_{s,n_1}(\f{\mu}{\mu_bq^{n_1}})^{n-s}
(\f{\mu_a}{\mu}q^{s+1};q)_{n-s} \cr}~~~~~~~~~~~~~n\ge s,
\ee
for some $s\in \zn$, implying
\be
\sum\limits_{n=s}^{\infty}\bar{\varphi}_{n,n_1}^*\varphi_{n,n_1}=
\bar{\varphi}_{s,n_1}^*\varphi_{s,n_1}\sum\limits_{n=s}^{\infty}1=\infty.
\ee
We add the further label $n_2$ to label the corresponding solutions.
Replacing this value in equations (4.34-35) we find
$$
\varphi_{n,n_1,n_2}=\cases{\varphi_{n_2,n_1,n_2}
(\f{\mu_bq^{n_1-n_2}}{\mu_a})^{n-n_2}
\f1{(q;q)_{n-n_2}} ~~~~~~~~~~if~~n\ge n_2\cr 0~~~~~~~~~~~~~~~~~~~~~~~~~
{}~~~~~~~~~~~~~~~~~~~~~~~~~~otherwise\cr}
$$
\be
\bar{\varphi}_{n,n_1,n_2}=\cases{
\bar{\varphi}_{n_2,n_1,n_2}(\f{\mu_bq^{n_1-n_2}}{\mu_a})^{n_2-n}
\f1{(q^{-1};q^{-1})_{n_2-n}} ~~~~~~~~~~
if~~n\le n_2\cr 0~~~~~~~~~~~~~~~~~~~~~~~~~~~~~~~~~~~~~~~~~~~~~~~~~~~~~
otherwise.\cr}
\ee
We see that $|\bar{\varphi}_{n_1,n_2}>\in\h_{\mu_a}$ $\forall n_1,n_2\in\zn$,
whereas $|\varphi_{n_1,n_2}>\in\h_{\mu_a}$ only if $n_1\ge n_2$ (otherwise
its norm is $\infty$; for $q>1$ the situation is exactly inverse).

Define
\be
\Vert n_2,n_1>:=(|\varphi_{n_1,n_2}>_a\ot|n_1>_b,
|\bar{\varphi}_{n_1,n_2}>_a\ot|n_1>_b)
\ee
and
\be
{\cal U}^{phys}:=Span_{\cn}\{\Vert n_2,n_1>~~~,~~~n_1-n_2\ge 0\}.
\ee
According to definition (4.28), the square norm  of $\Vert n_2,n_1>$ is
equal to $\bar{\varphi}_{n_2,n_1,n_2}^*\varphi_{n_2,n_1,n_2}$
In order to get a positive norm it is sufficient to take  the latter
quantity positive. We will choose
$\bar{\varphi}_{n_2,n_1,n_2}=1=\varphi_{n_2,n_1,n_2}$. Consequently,
\begin{prop}
The space of `` physical '' two particle states
${\cal U}^{phys}$ is a pre-Hilbert space with scalar product (4.28), and
${\cal B}_2:=\{\Vert n_2,n_1>~~,~~n_1-n_2\ge 0\}$ is an orthonormal basis,
such that
$$
P_1\Vert n_2,n_1>=\mu_bq^{n_1}\Vert n_2,n_1>
{}~~~~~~~~~~~~~~~~~P_2\Vert n_2,n_1>=\mu_aq^{n_2}\Vert n_2,n_1>
$$
\be
\La_1^{\pm 1}\Vert n_2,n_1>=\Vert n_2,n_1\pm 1>
{}~~~~~~~~~~~~~~~~~\La_2^{\pm 1}\Vert n_2,n_1>=\Vert n_2\pm 1,n_1>.
\ee
Moreover
\be
\cases{
|\varphi_{n_1,n_2}>=\sum\limits_{n=0}^{\infty}(\f{\mu_bq^{n_1-n_2}}{\mu_a})^n
\f 1{(q,q)_n}|n_2+n> \cr
|\bar{\varphi}_{n_1,n_2}>=\sum\limits_{n=0}^{\infty}(\f{\mu_bq^{n_1-n_2}}{\mu_a})^n
\f 1{(q^{-1},q^{-1})_n}|n_2-n>. \cr}
\ee
\end{prop}
{\bf Remark}. The spectrum of the total momentum $P_2$ is
$\{\mu_aq^n\}_{n\in\zn}$, the same as that of $p\ot \1$; the momentum scale
$\mu_a$ is the same as that of the first particle.
The eigenvectors
of $\r(P_2),\br(P_2)$ are {\it superpositions} of eigenvectors of
$p\ot\1$, i.e. a sort of interaction between the first and second
particle arises.

The name `` physical '' is due to the fact that,
as a consequence of its definition,
${\cal U}_{phys}$ is characterized by the physical condition that on
each eigenvector of $P_1,P_2$ (recall that $P_2$ represents the total
momentum of the two-particle system, whereas $P_1$ the momentum of the second
particle)
\be
|\!<P_2>\!|~>~|\!<P_1>\!|~~~~~~if~~\f{\mu_a}{\mu_b}>0;
\ee
compare with eq.'s (2.8).

Finally, note that
\be
\La_1^{-1}\Vert n,n>\not\in{\cal U}^{phys}
{}~~~~~~~~~~~~~~~~~\La_2\Vert n,n>\not\in{\cal U}^{phys},
\ee
since the corresponding unbarred series have infinite norm in
$\h_{\mu_a}\ot\h_{\mu_b}$.

\sect{Generalization to $n$ particles}

{}~~~Let us define for the sake of brevity
$$
P_i:=(\r_n(P_i),\br_n(P_i))
{}~~~~~~~~~~~~\La_i:=(\r_n(\La_i),\br_n(\La_i))
$$
\be
\cases{
\r_n(P_i):=\underbrace{\1\ot...\ot\1}_{(n-i)~~times}\ot\phi_i(p) \cr
\br_n(P_i):=\underbrace{\1\ot...\ot\1}_{(n-i)~~times}\ot\bphi_i(p) \cr}
\ee
$$
\cases{
\r_n(\La_i):=\underbrace{\1\ot...\ot\1}_{(n-i-1)~~times}\ot
\f{\1\ot\phi_i(p)}{\phi_{i+1}(p)}(p\ot p^{-1}\la\ot
\underbrace{\1\ot...\ot\1}_{(i-1)~~times})  \cr
\br_n(\La_i):=\underbrace{\1\ot...\ot\1}_{(n-i-1)~~times}\ot
\f{\bphi_{i+1}(p)}{\1\ot\bphi_i(p)}(p^{-1}\ot p\la\ot
\underbrace{\1\ot...\ot\1}_{(i-1)~~times})  \cr}~~~~~~~~~~if~~~n>i.
$$
\be
\cases{
\r_n(\La_n):=\phi_n(p)(\la p^{-1}\ot
\underbrace{\1\ot...\ot\1}_{(n-1)~~times} ) \cr
\br_n(\La_n):=\f 1{\bphi_n(p)}(\la p\ot
\underbrace{\1\ot...\ot\1}_{(n-1)~~times}  )\cr}
\ee
(($\phi_1(p)\equiv p$). We denote by $P_i^{-1},\La_i^{-1}$ the inverses
of $P_i,\La_i$. Then

\begin{prop}
$P_i,\La_i$, $~i,j=1,2,...,n$, have well-defined commutation relations
\be
[P_i,P_j]=0~~~~~~~[\La_i,\La_j]=0~~~~~~~~P_i\La_j=q^{\delta_i^j}\La_jP_i
\ee
and therefore form together with $P_i^{-1},\La_i^{-1}$ a set of generators of
$(\bigotimes^n\A)_d:=\{(\alpha,\gamma_n(\alpha))~~|~~\alpha\in\bigotimes^n\A\}$
, $\gamma_n:=\br_n\circ\r_n^{-1}$. Moreover definition (3.20) introduces a
complex conjugation $*_n$ which is compatible with the algebra relations (5.3):
\be
P_i^{*_n}=P_i~~~~~~~~~~~\La_i^{*_n}=\La^{-1}_i
\ee
\end{prop}

$Proof$ The proof of the commutation relations (5.3) is recursive. Assume
that they hold when $n=m$. The corresponding $\r$ images read:
\be
[\r_m(P_i),\r_m(P_j)]=0~~~~~~~~i,j=1,...,m
\ee
Let
\be
\phi_{m,l}:=(\underbrace{id\ot...\ot id}_{(m-l)~~times}\ot\phi\ot
\underbrace{id\ot...\ot id}_{(l-1)~~times})~~~~~~~~~~1\le l\le m;
\ee
now note that from definition (5.1) and equation (1.3) it follows
\be
\phi_{m,l}(\r_m(P_i))=
\cases{\r_{m+1}(P_i)~~~~~~~~~if~~~l> i \cr
       \r_{m+1}(P_{i+1})~~~~~~~~if~~~l\le i \cr}
\ee
Therefore applying the operators $\phi_{m,l}$ to both sides of
equations $(5.3)_1$ in the case $n=m$
we get the $\r$ images of equations  $(5.3)_1$ in the case
$n=m+1$. The remaining relations $(5.3)_2,(5.3)_3$ are most simply
proved by direct calculation, after noting that
\be
\left[\r_n(P_i),
\underbrace{\1\ot...\ot\1}_{(n-j-1)~~times}\ot p\ot p^{-1}\la\ot
\underbrace{\1\ot...\ot\1}_{(j-1)~~times}\right]_{q^{\delta_i^j}}=0.
\ee
The proof is similar for the $\br$-images. $\diamondsuit$

The procedure described in section 4 can be iterated in a
straightforward way to yield a $*_n$-representation of $(\bigotimes^n\A)_d$
on a Hilbert space
${\cal U}^{phys}_{\mu_n,...,\mu_1}\subset
(\bigotimes\limits_{i=1}^n\h_{\mu_{n-i+1}},\bigotimes\limits_{i=1}^n\h_
{\mu_{n-i+1}})$
which is the linear span of a set of vectors
${\cal B}_n:=\{\Vert n_n,...,n_1>,~~~n_{i+1}\le n_i\}$
where, for $i=1,2,...,n$,
\be
P_i\Vert n_n,...,n_1>=\mu_iq^{n_i}\Vert n_n,...,n_1>
{}~~~~~~~~~~~~~~~~~~~\La_i^{\pm 1}\Vert n_n,...,n_1>=
\Vert n_n,...,n_i\pm 1,...,n_1>;
\ee
moreover ${\cal B}_n$ is an orthonormal basis of
${\cal U}^{phys}_{\mu_n,...,\mu_1}$ w.r.t. the scalar product (4.28).

It goes as follows: assume that one has been found for $n=m$, and
that the mass scale of $P_m$ (the total momentum of $m$ particles)
is $\mu_m$ (i.e. the mass scale of the Hilbert space which is the first
tensor factor in $\bigotimes\limits_{i=1}^m\h_{\mu_{m-i+1}}$).
Then we introduce vectors $\Vert n_{m+1},n_m,...,n_1>$ by
replacing in formula (4.31) to $\mu_a,n_2$ respectively
$\mu_{m+1},n_{m+1}$, and to
the vector $|n_1>_b$ the vector $\r(\Vert n_m,n_{m-1},...,n_1>)$, and similarly
for $\br_{m+1}$, in other words we set
\be
\cases{
\r_{m+1}(\Vert n_{m+1},n_m,...,n_1>):=
|\varphi_{n_m,n_{m+1}}>_{m+1}\ot\r(\Vert n_m,...,n_1>)  \cr
\br_{m+1}(\Vert n_{m+1},n_m,...,n_1>):=
|\bar{\varphi}_{n_m,n_{m+1}}>_{m+1}\ot\br(\Vert n_m,...,n_1>) \cr}
{}~~~~~~~~~~\Vert n_m,...,n_1>\in {\cal B}_m
\ee
and
\be
\Vert n_{m+1},n_m,...,n_1>=\left(
\r_{m+1}(\Vert n_{m+1},n_m,...,n_1>),
\br_{m+1}(\Vert n_{m+1},n_m,...,n_1>)\right)
\ee
The set ${\cal B}_{m+1}:=\{\Vert n_{m+1},n_m,...,n_1>,~~~n_{i+1}\le n_i\}$
is a set of eigenvectors of $P_i$, $i=1,2,...,m+1$
with eigenvalues $\mu_iq^{n_i}$, orthonormal w.r.t.
the product (4.28) where in the RHS $\h_a$ stands for the space
$\h_{\mu_{m+1}}$ and $\h_b$ for the space ${\cal U}_{\mu_m,...,\mu_1}$.
Moreover, a glance at definition (3.20) (of
$*_{m+1}$) will convince the reader that $*_{m+1}$ on
$P_i,\La_i$ $i=1,2,...,m+1$ acts as hermitean conjugation w.r.t.
the scalar product (4.28). Finally
\be
P_i\Vert n_{m+1},...,n_1>=\mu_iq^{n_i}\Vert n_{m+1},...,n_1>
{}~~~~~~~~~~~~~~~~~~~\La_i^{\pm 1}\Vert n_{m+1},...,n_1>=
\Vert n_{m+1},...,n_i\pm 1,...,n_1>,
\ee
as claimed.

\sect{Configuration space realization}

{}~~~It is possible to realize the Hilbert
space $\h_{\mu}$ as a pair of subspaces of $Fun(\rn_q)$ and the algebra
$\A$ as a subalgebra of the differential algebra $Dif\!f(\rn_q)$ on
$\rn_q$, in other words in terms of q-differential operators acting
on $Fun(\rn_q)$. The $*$ is realized as the one of $Dif\!f(\rn_q)$, and the
scalar product of $\h_{\mu}$ through a q-integral \cite{kempf} .
The procedure is a word by word translation of the configuration space
realization of the harmonic oscillator \cite{fio} and of the free
particle \cite{fioeu} on $\rn_q^N$; this is the source of inspiration
for the main idea of the present work.
Moreover, the construction of multi-particle systems can be
done as above and is best understood in terms of braided configuration
spaces.

Let $Fun(\rn_q)$ be the unital algebra generated by $x$,
$Dif\!f(\rn_q)$ the unital algebra generated by $\p,x$ with derivation
relations
\be
\p x=1+qx\p.
\ee
The dilaton $\tla$ is defined by
\be
\tla:=[\p,x]=1+(q-1)x\p~~~~~~\Rightarrow~~~~~~~~\tla x=qx\tla,
{}~~~~~~~\tla\p=q^{-1}\p\tla.
\ee
If we introduce a complex conjugation $\s$ by requiring that $x$ is real then,
\be
x^{\s}=x,~~~~~~~~~\p^{\s}=-q^{-1}\bp ~~~~~~~~~\Rightarrow
\tla^{\s}=q^{-1}\tla^{-1},
\ee
where $\bp$ is the barred derivative
\be
\bp:=\tla^{-1}\p,~~~~~~~~~~~~\bp x=1+q^{-1}x\bp.
\ee
Contrary to the abstract treatment of section 4, here it
is convenient also in the case of a one-particle system to give a pair
of realizations both of $\A$ and of $\h_{\mu}$, we will call them
$\tilde{\r_1}(\A),\tilde{\r_1}(\h_{\mu})$ and
$\tilde{\br_1}(\A),\tilde{\br_1}(\h_{\mu})$ respectively. We start
from
\be
\cases{\tilde{\r_1}(p)=i\p \cr
\tilde{\br_1}(p)=i\bp \cr}~~~~~~~~~~~\cases{\tilde{\r_1}(\la^{\pm 1})=\tla \cr
\tilde{\br_1}(\la^{\pm 1})=q^{\pm 1}\tla^{\pm 1} \cr}.
\ee
Setting $\A'_d:=\{(\tilde{\r_1}(a),\tilde{\br_1}(a))~~|~~a\in\A\}\ap \A$,
one can realize the $*$-involution of $\A$ on $\A'_d$ by setting
\be
(\tilde{\r_1}(a^*),\tilde{\br_1}(a^*))=\tau\circ(\s,\s)[
(\tilde{\r_1}(a),\tilde{\br_1}(a))].
\ee
The $\tilde{\r_1},\tilde{\br_1}$ images of the vectors $|n>$ are determined as
functions of $x$ which satisfy the $\tilde{\r_1},\tilde{\br_1}$ image of the
eigenvalue equations
$p|n>=\mu q^n|n>$ and of the `` step '' equation
$\La^{\pm 1}|n>=|n\pm 1>$:
\be
\cases{
\tilde{\r_1}(|n>):=e_q[-i\mu q^nx]  \cr \tilde{\br_1}(|n>):=q^ne_{q^{-1}}[-i\mu
q^nx] \cr}
\ee
where
\be
e_q[Z]:=\sum\limits_{l=0}^{\infty}\f{Z^l}{l_q!}~~~~~~~~~~~l_q:=
\f{1-q^l}{1-q};
\ee
$\tilde{\r_1},\tilde{\br_1}$ are extended on $\A$ as algebra isomorphisms, on
$\h_{\mu}$ as vector space isomorphisms. Therefore to each $|\psi>\in\h_{\mu}$
there will correspond two functions $\psi(x):=\tilde{\r_1}(|\psi>)$ and
$\bar{\psi}(x):=\tilde{\br_1}(|\psi>)$. The scalar product in
$\h_{\mu}$ will be realized through algebraic q-integration \cite{kempf,jack}
\be
<\psi'|\psi>=\int d_qx[\bar{\psi}'(x)^{\s}\psi(x)+(\psi')^{\s}(x)\bar{\psi}(x)]
\ee
after fixing its normalization properly. Due to the properties of
q-Stokes theorem of q-integration
\be
\int d_qx \p f(x)=0=\int d_qx \bp f(x),
\ee
definition (6.9) is compatible with
hermitean conjugation of the $*$-representation
of $\A$ on $\h_{\mu}$.
All the treatment done in the abstract case applies here too, provided we
modify the definitions of $(\bigotimes^n\A)_d$ into
$(\bigotimes^n\A)'_d:=(\bigotimes^n \tilde{\r_1},\bigotimes^n \tilde{\br_1})
[(\bigotimes^n\A)_d]$.
One can easily verify that
\be
\cases{
\tilde{\r_1}(|\varphi_{n_1,n_2}>)(x)=:\varphi_{n_1,n_2}(x)=\sum\limits_{l=0}^
{\infty}[i\mu_aq^{n_2}(q-1)x]^l
\f{(\f{q^{n_1-n_2}\mu_b}{\mu_a};q)_l}{(q;q)_l} \cr
\tilde{\r_1}(|\bar{\varphi}_{n_1,n_2}>)(x)=:\bar{\varphi}_{n_1,n_2}=
\sum\limits_{l=0}^{\infty}[i\mu_aq^{n_2}(q^{-1}-1)x]^l
\f{(\f{q^{n_1-n_2-1}\mu_b}{\mu_a};q^{-1})_l}{(q^{-1};q^{-1})_l}, \cr}
\ee
by solving the corresponding eigenvalue equations in $Fun(\rn_q)$.

\sect{Classical limit}

In order that the toy-model considered in this work can be considered as
physically realistic, it should describe a system of one or more free
particles on a continuous line in the limit
(understood in some reasonable sense) $q\rightarrow 1^-$.

Let us denote by $|n>_q$ the orthonormal basis vectors of $\h_{\mu}$
for a given $q<1$, $p|n>_q=\mu q^n|n>_q$. For any fixed $n$
$\lim_{q\rightarrow 1^-}\mu q^n=\mu$, i.e. all eigenvalues collapse
(but $not$ uniformly) to $\mu$,
therefore the set $\{\lim_{q\rightarrow 1^-}|n>_q\}_{n\in\zn}$
is no generalized basis of the representation space of the undeformed
1-dimensional translation+dilatation group, and,
as already noticed in ref. \cite{fioeu}, the limit $q\rightarrow 1^-$
cannot be understood a naive sense.

A more adequate notion of such a limit seems the following
(see also Ref. \cite{fioeu}). For each
nonnormalizable eigenvector (i.e. distribution) $|\pi>_c$ of the
classical momentum $p$, $p|\pi>_c=\pi|\pi>_c$, there exists a function
$\tilde{n}(\pi,q)$ such that $|\pi>_c=\lim_{q\rightarrow 1^-}|n(\pi,q)>_q$,
in the sense of convergence of the eigenvalues.
Such a problem admits the following solution
\be
\tilde{n}(\pi,q)=[log_q(\f{\pi}{\mu})],
\ee
where $[a]$ denotes the integral part of $a\in\rn$. In fact
the eigenvalue of $p$ on $|\tilde{n}(\pi,q)>_q$ goes to $\pi$ in the
limit $q\rightarrow 1^-$, and the expression
$\lim_{q\rightarrow 1^-}|n(\pi,q)>_q$ could be given a sense as a distribution
on ${\cal S}(\rn)$. Therefore we can say that the toy-model is
appropriate for an approximate description of $one$ ordinary free particle,
provided we choose $q$ sufficiently close to 1.

When considering $many$-particle systems, we have seen that an unavoidable
sort of interaction arises as a consequence of the inthrinsic braided
character of the total momentum. Let us stick for simplicity to the case
of two particles. In section 4 we have seen that the eigenvectors
$\Vert n_1,n_2>_q$ of $P_1,P_2$ is the tensor product of
two one-particle states,
the second is a free one-particle state $|n_1>_q$, whereas the first is a
superposition of different free one-particle states; in the $\r$ realization
it is denoted by $|\varphi_{n_1,n_2}>_q$. In order that the two-particle
states $\Vert n_1,n_2>_q$ are realistic descriptions of the classical free
two-particle states $\Vert \pi_1,\pi_2>_c$ (defined by
$P_i\Vert \pi_1,\pi_2>_c=\pi_i\Vert \pi_1,\pi_2>_c$, $\pi_2>\pi_1$),
we should check that in the unbarred representation in the limit
$q\rightarrow 1^-$ the superposition $|\varphi_{\tilde{n}_1,\tilde{n}_2}>_q$
($\tilde{n}_i(q,\pi_i):=[log_q(\f{\pi_i}{\mu_i})]$), becomes a
distribution with well-defined momentum $\pi_2-\pi_1$, i.e. that the
interaction effect disappears. Let us check that this is the case.
Let $\pi:=\f{\pi_1}{\pi_2}$ ($\pi<1$). Then
\be
|\varphi_{\tilde{n}_1,\tilde{n}_2}>_q=
\sum\limits_{n=0}^{\infty}c_n|n+n_2(q,\pi_2)>~~~~~~~~~~~~~~~~~~~
c_n:=\f{(\pi r(q,\pi))^n}{(q;q)_n}
\ee
where the rest $r(q,\pi):=(\f{\mu_bq^{\tilde{n}_1-\tilde{n}_2}}{\pi\mu_a})$
satisfies the relation $q^2\le r(q,\pi)<1$
($\lim_{q\rightarrow 1^-}r=1$). The function $(\pi r(q,\pi))^n$ decreases
with $n$, the function $\f 1{(q;q)_n}$ increases with $n$, therefore there
is a value $n=\bar{n}(q,\pi)$ for which $c_n$ is maximum. To determine
$\bar{n}$ (up to an error of 1) we consider $n$ as a continuous variable and
look for the solution of the equation $\f{\p}{\p n}c_n=0$. We are actually
interested in the asymptotic solution in the limit $q\rightarrow 1^-$.
In order to properly differentiate  the function in the denominator
we note that $\f 1{(q,q)_n}:=\f{(q^{n+1};q)_{\infty}}{(q;q)_{\infty}}$.
We find
\be
0\equiv\f{\p}{\p n}c_n=c_n[ln(\pi)-ln(q)g(n,q)],
\ee
where
\be
g(n,q):=\sum\limits_{k=0}^{\infty}\f{q^{n+k+1}}{1-q^{n+k+1}}=
\sum\limits_{k=0}^{\infty}q^{n+k+1}\sum\limits_{l=0}^{\infty}q^{(n+k+1)l}
=\sum\limits_{l=0}^{\infty}q^{(n+1)(l+1)}\sum\limits_{k=0}^{\infty}q^{k(l+1)}=
\sum\limits_{l=0}^{\infty}\f{q^{(n+1)(l+1)}}{1-q^{l+1}}.
\ee
Hence
\be
ln(q)g(n,q)\stackrel{q\rightarrow 1^-}{\Large \ap}
=-\sum\limits_{l=0}^{\infty}\f{q^{(n+1)(l+1)}}{(l+1)_q}
\stackrel{q\rightarrow 1^-}{\Large \ap} ln(1-q^{n+1})
\ee
since the q-deformed integer $l_q:=\f{q^l-1}{q-1}$ goes to $l\in\zn$
in the limit $q\rightarrow 1^-$; therefore the searched
solution $\bar{n}(q,\pi)$ is
such that $q^{\bar{n}}\stackrel{q\rightarrow 1^-}{\Large \ap}1-\pi$. This
implies that in the superposition $|\varphi_{\tilde{n}_1,\tilde{n}_2}>_q$
the eigenvector with the eigenvalue $p=\pi_2(1-\pi)=\pi_2-\pi_1$ is enhanced.
Furthermore, by setting $f(u):=c_n\vert_{n=[log_q(1-u)]}$ one could
show that $f(u)$ is more and more peaked around $u=\pi$
in the limit $q\rightarrow 1^-$. This means that in that limit
$|\varphi_{\tilde{n}_1,\tilde{n}_2}>_q$ goes to the distribution describing
a free particle with momentum $\pi_2-\pi_1$ on $\rn$, as claimed.

\section{Appendix}

In this appendix we clarify in which sense  for a
$n$-particle system the set $\{P_1,P_2,...,P_n\}$
considered in section 5 is the only admissible complete set of
commuting observables including the total momentum $\Phi_n(p)$.
For the sake of brevity we stick to the case $n=2$.

Let
\be
a:=\phi(p),~~~~~~~b:=\1\ot p,~~~~~~~c:=p\ot \la,~~~~~~~d:=\la\ot\la.
\ee
It is immediate to realize that $a,b,c,d$ and their inverses are
a complete set of independent generators of $\A\ot\A$, and
$[a,b]=0=[a,c]$. The unbarred image of any observable $V$ independent of
$\Phi(p)$ and commuting with it
is therefore a function of $b,c$ only. The subalgebra
${\cal V}$ generated by $b,c$ can be decomposed as
${\cal V}=\bigoplus\limits_{n\in\zn}({\cal V}_1)^n$, where
${\cal V}_1$ is its component satisfying the relation
$[{\cal V}_1,d]_q=0$, i.e. with natural dimension 1.
Thus we can search $v=\r(V)$ within ${\cal V}_1$ without loss of
generality. ${\cal V}_1$ is a subspace spanned by $c^{-j}b^{j+1}$;
therefore $v$ has to be a combination (which we will assume to be finite)
\be
v=\sum\limits_{j=J_0}^{J_1} v_jc^{-j}b^{j+1}\in{\cal V}_1.
\ee
The pair $V:=(v,v^{*\ot *})$ is a real element of $(\A\ot\A)_d$;
we will say that $V$ is admissible (and consequently can be chosen
as an observable, beside $\Phi(p)$)
if the eigenvectors $|\psi>$ of $\phi(p),v$ are elements of
(i.e. normalizable in) $\h_{\mu_a}\ot\h_{\mu_b}$. We ask now for
which choice of the coefficients $v_i$ $V$ is admissible.

\begin{prop}
$V:=(v,v^*)$ is admissible only if $v\propto b$; in other words
$P_1$ is the only commuting observable which we can add to $P_2=\Phi(p)$.
\end{prop}
$Proof$. Reasoning as in section 4, it is easy to check that, independently
of the choice of $v$, the spectrum of $a$ is $\{\mu_aq^n\}_{n\in\zn}$,
and
\be
a|\psi_{n_2}>=\mu_a q^{n_2}|\psi_{n_2}>~~~~~\Rightarrow~~~~~~
|\psi_{n_2}>:=\sum\limits_{m=0}^{\infty}
c_{n_2,m}\left[\sum\limits_{n=0}^{\infty}
\f{(\f{\mu_b}{\mu_a}q^m)^n}{(q;q)_n}|n+n_2>\right]\ot|m+n_2>.
\ee
The requirement of normalizability in $\h_{\mu_a}\ot\h_{\mu_b}$ of
$|\psi_{n_2}>$ implies
\be
\sum\limits_{m=0}^{\infty}|c_{n_2,m}|^2~ <~
\sum\limits_{m=0}^{\infty}|c_{n_2,m}|^2d_m=<\psi_{n_2}||\psi_{n_2}>_{a\ot b}
<\infty,~~~~~~~~~~~(here~~d_m:=\sum\limits_{n=0}^{\infty}
\left[\f{(\f{\mu_b}{\mu_a}q^m)^n}{(q;q)_n}\right]^2>1);
\ee
consequently there exists a constant $C>0$ such that
$\vert\f{c_{n_2,m+h}}{c_{n_2,m}}\vert\le C$ if $h>0$, $m$ is large enough
and $c_{n_2,m+h}\neq 0\neq c_{n_2,m}$. Now assume that $|\psi_{n_2}>$
is also an eigenvector of $v$ with eigenvalue $\mu$. By easy computations
one shows that
the eigenvalue equation $(v-\mu)|\psi_{n_2}>=0$ amounts to
\be
\theta(k-J_1)
\sum\limits_{j=\max(J_0,-k)}^{J_1}\hat v_jc_{n_2,j+k}q^{k(j+1)+n_2}=
\theta(k)\f{\mu}{\mu_b}c_{n_2,k},
\ee
where $\hat v_j:=v_j(\f{\mu_b}{\mu_a})^jq^{j(j+1)}$ and
$\theta(s)=\cases{0~~~if~~s<0 \cr 1~~~if s\ge 1\cr}$.

It is easy to show that if $J_1\neq 0$ then $c_{n_2,k}=0$ $\forall k\ge 0$.
In fact, if $J_1<0$, then choosing first $k=0$, then $k=1$, $k=2$ etc,
we iteratively
show that the LHS in equation $(8.10)$ vanishes, and so does $c_{n_2,k}$.
Similarly, if $J_1>0$, then choosing first $k=-J_1$, then $k=-J_1+1$,
$k=-J_1+2$ etc, we iteratively
show that the RHS in equation $(8.10)$ vanishes, and so does $c_{n_2,k}$.
Therefore a nontrivial solution is possible only if $J_0\le J_1=0$.
Under these circumstances one can rewrite equation (8.10) as follows:
\be
c_{n_2,k}\left(\f{\mu}{\mu_b}-q^{k+n_2}\hat v_0\right)=
\sum\limits_{j=J_0}^{-1}\hat v_jc_{n_2,j+k}q^{k(j+1)+n_2}
\ee
Unless $J_0=0$, the RHS is different from zero and
in the limit $k\rightarrow \infty$ its dominant contribution comes
from $j=J_0$, because of the diverging factor $q^{k(j+1)}$ and the
bound $\vert\f{c_{n_2,k+j+h}}{c_{n_2,k+j}}\vert\le C$:
\be
c_{n_2,k}\left(\f{\mu}{\mu_b}-q^{k+n_2}\hat v_0\right)
\stackrel{q\rightarrow 1^-}{\Large \ap}
\hat v_{J_0}c_{n_2,J_0+k}q^{k(J_0+1)+n_2};
\ee
consequently, $|c_{n_2,k}|$ does not converge to zero because
of the divergent factor $q^{k(J_0+1)}$, against the hypothesis.
Therefore, it must be $J_0=J_1=0$, namely,
we have proved the claim . $\diamondsuit$

\end{document}